\documentclass[twocolumn,showpacs,preprintnumbers,amsmath,amssymb]{revtex4}
\usepackage{graphicx}% Include figure files
\usepackage{dcolumn}% Align table columns on decimal point
\usepackage{bm}% bold math

\begin{document}

\preprint{}

\title{
Spin bottleneck in resonant tunneling through double quantum dots\\ with different Zeeman splittings
}% Force line breaks with \\

\author{S. M. Huang$^{1,2}$, Y. Tokura$^{3,4}$, H. Akimoto$^1$, K. Kono$^1$, J. J. Lin$^{2}$, S. Tarucha$^{4,5}$, and K. Ono\footnote{ E-mail address: k-ono@riken.jp}$^{1,4}$ }
\affiliation{$^1$Low Temperature Physics Laboratory, RIKEN, Wako-shi, Saitama 351-0198, Japan\\
$^2$Institute of Physics, National Chiao Tung University, Hsinchu 30010, Taiwan\\
$^3$NTT Basic Research Laboratories, NTT Corporation, Atsugi-shi, Kanagawa 243-0198, Japan\\
$^4$Quantum Spin Information Project, ICORP-JST, Atsugi-shi, Kanagawa 243-0198, Japan\\
$^5$Department of Applied Physics, University of Tokyo, Bunkyo-ku, Tokyo 113-8656, Japan}

\date{\today}% It is always \today, today, but any date may be explicitly specified

\begin{abstract}%no more than 600 characters, including spaces

We investigated the electron transport property of the InGaAs/GaAs double quantum dots, the electron g-factors of which are different from each other. We found that in a magnetic field, the resonant tunneling is suppressed even if one of the Zeeman sublevels is aligned. This is because the other misaligned Zeeman sublevels limit the total current. A finite broadening of the misaligned sublevel partially relieves this bottleneck effect, and the maximum current is reached when interdot detuning is half the Zeeman energy difference.

\end{abstract}

\pacs{ 73.63.Kv, 72.25.Mk, 73.23.Hk}
% Classification Scheme. 73.63.Kv Quantum dots, 72.25.Mk Spin transport through interfaces, 73.23.Hk Coulomb blockade; single-electron tunneling
%\keywords{Suggested keywords}%Use showkeys class option if keyword display desired
\maketitle

%Introduction
Electron g-factors in III-V semiconductor heterostructures can be tuned by changing the alloy ratio and thickness of each quantum well. Novel spin-related physics in such a g-factor-tuned system has attracted considerable interest in the past decade \cite{Spintronics}. Electrical tunings of electron g-factors has been demonstrated in single and coupled double quantum wells. The g-factor tunings are accomplished by changing the position of the electron wave functions that is spatially delocalized over the regions with different g-factors \cite{Jiang, Salis, Poggio}. The g-factor tuning used above is a powerful tool for the coherent manipulation of the electron spins, however these works were done for an ensemble of spins in quantum wells. To realize the spin based quantum information devices, which is a one of the most ambitious applications of semiconductor spintronics, the coherent manipulation in a single spin level is necessary  \cite{Loss}. The quantum dot is known as a system where a number of electron as well as their spin state can be well defined and easily controlled \cite{Hanson}. An application of such g-factor tuning to quantum dot systems offers a novel candidate for future electric devices for coherent spin manipulation. Tuning the Zeeman splitting $g_i\mu_BB$ of each spin in a quantum dot array has been proposed for individual addressing of spin qubits and fast gate operation between two qubits, where $g_i$ ($i = 1, 2, \cdots$) is different for each dot \cite{Loss, Coish}. Selective addressing has been demonstrated so far in a double quantum dot made from spatially homogeneous g-factors with an additional micro-ferromagnet nearby. The magnet creates a $\sim$10mT field difference of external magnetic field $B$ in each dot \cite{Micromagnet}. Quantum dots with different $g_i$ can offer a much larger Zeeman energy difference in the same external field with smaller spatial separation.

In this letter, we report a novel behavior of a single spin in a g-factor-tuned double dot, where Zeeman splittings differ greatly from each other in a magnetic field. We investigate this system in a simple, well-defined regime where the total electron number $N$ in the double dot is zero or one. Accompanying the theoretical calculation, we reveal that the resonant tunneling via two Zeeman-split levels is suppressed even if one pair of Zeeman-split levels is aligned. This novel spin bottleneck effect is partially relieved by a finite broadening of zero-dimensional states owing to interdot and/or dot-electrode tunnel couplings. As a result of competition between the bottleneck and the level broadening effect, the current is maximum in the configuration where interdot level detuning is set to \emph{half} the Zeeman energy difference.

%Sample and measurement
Vertical double quantum dots with different g-factors are formed in a submicron-scale pillar of a triple barrier structure with a surrounding Ti/Au Schottky gate, as schematically shown in left inset of Fig. 1. The  structure consists of seven layer structure including triple barriers and comprises, from top to bottom (or from left to right in the schematic potential energy landscape in the right inset of Fig. 1),
a gradiently n-doped Al$_{0.05}$Ga$_{0.95}$As source electrode,
7-nm-thick Al$_{0.30}$Ga$_{0.70}$As barrier,
7.5-nm-thick In$_{0.04}$Ga$_{0.96}$As well,
6.5-nm-thick Al$_{0.30}$Ga$_{0.70}$As center barrier,
10-nm-thick GaAs well,
7-nm-thick Al$_{0.30}$Ga$_{0.70}$As barrier,
and gradiently n-doped Al$_{0.05}$Ga$_{0.95}$As drain electrode.
The fabrication prosedure is the same with the previous vertical double dots \cite{Vertical double dot}. Measurements were performed in a dilution refrigerator at an effective electron temperature of $\sim$0.1$~$K and in magnetic fields of up to 15$~$T applied perpendicular to the wells \cite{In plane field}.

%Figure large scale diamonds, sample & potential schematic
\begin{figure}
 \includegraphics[scale=0.51]{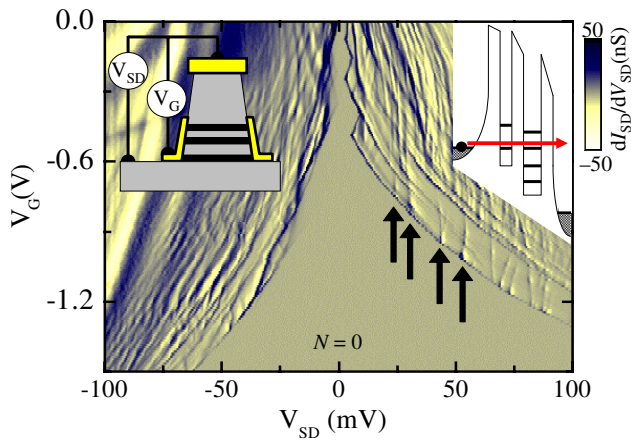}
 \caption{\label{fig1} $dI_{SD}/dV_{SD}$ plot of the sample measured at 0$~$T. The arrows mark current peaks due to resonant tunneling where the ground state of left dots is aligned to one of the excited states of the right dot, as shown in the right inset. Left inset: schematic of a gated vertical double quantum dot structure with different g-factors for each dot. The two quantum dots are made of InGaAs and of GaAs layers. We apply $V_{SD}$ to the substrate-side (bottom) electrode and measure the drain current from the top electrode. Right inset: potential energy landscape for the resonant tunneling condition.}
\end{figure}

%Basics of 0D-0D resonant tunneling
Figure 1 shows the differential conductance, $dI_{SD}/dV_{SD}$, plotted as a function of source-drain voltage $V_{SD}$ and gate voltage $V_{G}$ in zero magnetic field \cite{Asysmetry}. Current steps are recognized as dark blue lines. In the positive $V_{SD}$ region, near the current threshold, several current {\it peaks} (not steps) appear, as marked by arrows. These peaks are due to the resonant tunneling through the ground state of the left dot and an excited state of the right dot, as shown in the right inset. These current peak lines run nearly parallel to the $V_{G}$ axis because the side gate capacitively couples to the two dots almost equally, and the alignment of the two dot levels is nearly maintained against $V_{G}$. These behaviors have also been observed in vertical double dots with the same quantum well layers \cite{0D0D}. In the first Coulomb staircase defined as the $N=0$ threshold line and the neighboring parallel line, the current is carried by three charge configuration ($N_{1}$, $N_{2}$) = (0,0), (1,0), and (0,1), where $N_{i}$ ($i=1,2$), is the number of electron in each dot. In this configuration, the situation is simple enough to avoid electron-electron interactions in each dot.

%Figure Main experimental results
\begin{figure}
 \includegraphics[scale=0.46]{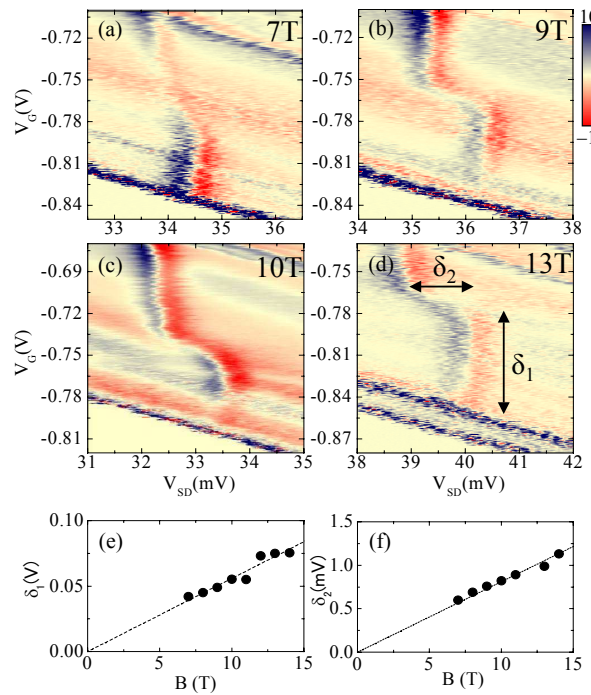}
 \caption{
\label{fig2}(a)-(d) Differential conductance, $dI_{SD}/dV_{SD}$, as a function of source-drain voltage and gate voltage under several different magnetic fields, showing increasing kink structure characterized by $\delta_1$ and $\delta_2$. Magnetic field dependences of (e) $\delta_1$ and (f)$\delta_2$.}
\end{figure}

%Experimental facts, extracting delta1 and delta 2
Application of the perpendicular magnetic field change the energies of excited orbital states of the dots, and shift $V_{SD}$ at the resonant tunneling peak lines \cite{0D0D}. Hereafter, we concentrate on the current peak line found around $V_{SD}\sim$35$~$mV and carry out detailed measurements under various magnetic fields. Figures 2(a)-(d) show the $dI_{DS}/dV_{SD}$ plots for several magnetic fields. The current peak line, which is recognized as adjacent blue and red lines, have a clear kink structure. In vertical double dots with the same quantum well layers, current peak lines are always straight in any magnetic field and the kink structure has never been observed. The kink is characterized by two values, $\delta_1$ and $\delta_2$, as marked in Fig. 2(d). Both $\delta_1$ and $\delta_2$ linearly increase with magnetic field, as shown in Figs. 2 (e) and (f). With low magnetic fields the peak line becomes straight asymptotically. Similar kink structures are found in all other current peak lines in various positive $V_{SD}$'s at high magnetic fields, as ling as that the width of the peak is narrow enough to resolve kinks.

%Figure schematic
\begin{figure}
 \includegraphics[scale=0.45]{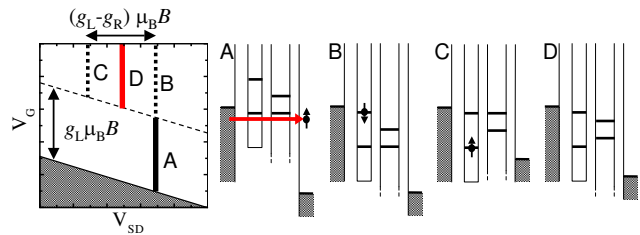}
 \caption{
\label{fig3} Schematic diagram of resonant tunneling peak line for Zeeman mismatched system, and characteristic potential landscapes A-D.}
\end{figure}

%Explanation with schematic figure, Spin bottleneck by Zeeman mismatch
In a double quantum dots with the same g-factors, both Zeeman sublevels are aligned at the same time at a certain $V_{SD}$, regardless of  the magnetic field \cite{0D0D}. In double quantum dots with different g-factors, however, the alignment of each Zeeman sublevel is achieved at its own $V_{SD}$, as schematically shown in Fig. 3, in accordance with the conditions of the measurements. We assign four particular conditions labeled A to D in the magnetic field. Under condition A, the aligned levels for up-spin for both dots is in the transport window and the resonant tunneling current is carried out by up-spin electrons. By increasing $V_G$, {\it i.e.}, by lowering the energy level of both dots, the down-spin Zeeman sublevel in the left dot comes within the transport window under condition B. Under this condition, although even the up-spin states are aligned and the resonant tunneling channel exists, the resonant tunneling is Coulomb blocked once the down-spin state in the left dot is occupied. We name this suppression process the spin bottleneck. Similarly, under condition C, the occupation of the up-spin in the left dot prohibits the subsequent tunneling even though the aligned down-spin channel exists. Thus under both conditions B and C, the bottleneck channel is occupied and there is no steady-state resonant current. It should be stressed that this bottleneck effect is similar to the Pauli spin blockade \cite{OnoSB, OnoNUC, Baugh}, since the stochastic single electron occupation of the bottleneck channel ultimately leads to a blockade in both cases.

The bottleneck can be lifted by level broadening, which is induced by finite tunnel couplings among the dots and the electrodes. The broadening couples the misaligned Zeeman sublevels and provides an escape path for the electron in the bottleneck channel, and relieves the bottleneck effect. The electron transport is carried out within the competition between the bottleneck and the escape effects. While smaller detuning in the resonant tunneling channel can carry large current, smaller detuning of the bottleneck channel is required to avoid the bottleneck. As a result of the compromise under the intermediate detuning condition D, not under B or C, the maximum current is expected.

%Figure Tokura calculation
\begin{figure}
 \includegraphics[scale=0.46]{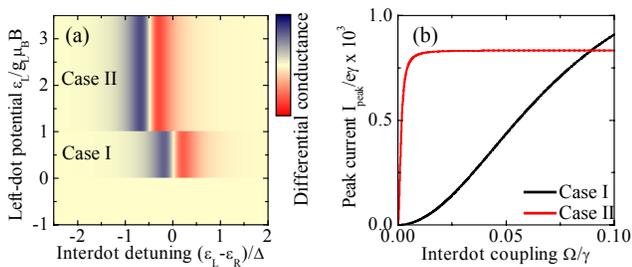}
 \caption{
\label{fig4}Calculated current by the Bloch equation method. (a) Differential conductance plotted as a function of interdot detuning (normalized by Zeeman energy difference $\Delta=|g_{L}-g_{R}|\mu_{B}B$) and left-dot potential (normalized by left-dot Zeeman splitting energy $|g_{L}|\mu_{B}B$), where $2.5\Delta=10\Omega=1000\gamma_{L}=100\gamma_{R}=\gamma$. Transport condition is categorized as two cases, I and II. We neglect a finite transition region between I and II due to small $\gamma_{L}$ here.(b) Peak current in cases I and II as a function of normalized interdot coupling $\Omega/\gamma$. Other parameters are the same as in (a).}
\end{figure}

%Tokura calculation and compare with measurement
This scenario is supported by the theoretical analysis. We evaluated the resonant tunneling current through two dots with different Zeeman splittings by the Bloch equation method \cite{Tokura, Stoof}. The quantum dot states are described by five bases, $|0,0\rangle, |\uparrow,0\rangle, |\downarrow,0\rangle, |0,\uparrow\rangle,$ and $|0,\downarrow\rangle$. Each dot has own g-factor, $g_L$ and $ g_R$ and each level has a Zeeman sublevel and their difference is $\Delta=|g_{L}-g_{R}|\mu_{B}B$ (hereafter, we assume $g_{L}<g_{R}<0$). In the calculation, we neglect the cotunneling and tunneling processes with phonon absorption/emission and assume zero temperature. We fix $V_{SD}$ to be large enough so that the right Fermi level is far below the right-dot level $\epsilon_{R}$. Figure 4(a) shows the differential conductance plotted as a function of $\epsilon_{L}-\epsilon_{R}$ and $\epsilon_{L}$. In the figure we set the origin of $\epsilon_{L}-\epsilon_{R}$ at the position where the lower Zeeman (up-spin) sublevels are aligned and $\epsilon_{L}$ is measured from the Fermi level of the left electrode. There are two conditions labeled case I and case II. In case I, only the up-spin Zeeman sublevel of the left dot is within the transport window (corresponds to scheme A in Fig. 3). In case II, both the up-spin and down-spin levels of the left dot are within the transport window (schemes B-D in Fig. 3). In both cases, we obtain a current with a single peak of Lorentzian shape. In case I, the current has a peak at zero interdot detuning. In case II, the maximum current is shifted in the negative detuning direction by the amount of $\Delta/2$. These peak positions in both cases I and II are found to be independent of parameters such as the coupling of the left (right) dot to its neighboring electrode, $\gamma_{L(R)}$, and interdot coupling $\Omega$. Note that $\epsilon_{L}-\epsilon_{R}$ and $\epsilon_{L}$ roughly correspond to $V_{SD}$ and $V_{G}$ in the measurement and $\delta_1$ and $\delta_2$ correspond to $|g_{L}|\mu_{B}B$ and $\Delta/2$, respectively. Thus the clear kink structure in the current peak line in Fig. 4(a) is similar to those in Figs. 2(a)-(d). Peak currents are plotted in Fig. 4(b) as a function of $\Omega$. The peak current in case II vanishes for small $\Omega$. In contrast, the two peak currents take similar values for large $\Omega$, where the dot-electrode couplings limit the current. In Figs. 2(a)-(d), the current peak heights are almost the same ($\sim$10$~$pA), both below and above the kink structure. This corresponds to the large $\Omega$. Indeed, our previous studies on vertical double dots with similar barrier thicknesses indicate $\Omega\sim$0.1$~$meV and $\gamma_{L(R)}\lesssim0.01$meV \cite{Vertical double dot}. In this sample, $\gamma_{L}$ is much smaller than $\gamma_{R}$ \cite{Asysmetry}. In Fig. 2 (a)-(d), transition between the regions corresponds to the case I and II is not as abrupt as in Fig. 4 (a) due to the finite temperature. It should be noted that, although this calculation reproduces the shift of the peak and the peak height well, the calculated peak width is always larger than the peak shift $\Delta/2$, whereas in the measurement, $\Delta/2$ seems to be twice as large as the peak width at 13$~$T. This discrepancy might be solved if we include the cotunneling effect.

%Figure L1-L4&microwave
\begin{figure}
 \includegraphics[scale=0.46]{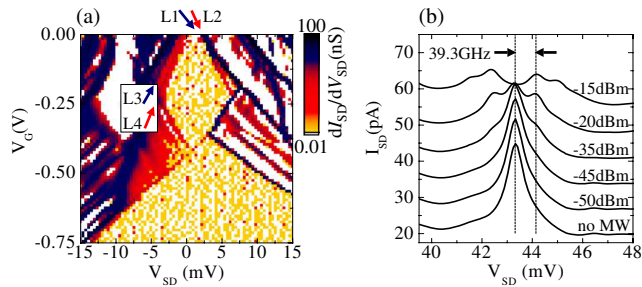}
 \caption{
\label{fig5}(a) $dI_{SD}/dV_{SD}$ plotted in logarithmic color scale in low-$V_{SD}$ region measured at 12$~$T. Four current threshold lines of the $N=1$ Coulomb diamond are marked as L1-L4. The slopes of these lines gives the voltage drop ratio of the triple barrier, $\alpha_i$ ($i=1, 2, 3$), for the low-$V_{SD}$ region as 0.27, 0.19, and 0.54.
(b) Resonant current peak under 39.3$~$GHz microwave application to $V_{SD}$. Interdot microwave-assisted tunneling is clearly apparent. Measurements were performed at 1.5$~$K using a different sample with the same triple barrier structure wafer. Distance between the main and satellite peaks was used to estimate the interdot voltage drop ratio $\alpha_2$ in high-$V_{SD}$ regions as 0.193. The microwave powers are labeled for each trace. Each trace is shifted by a constant value.}
\end{figure}

%from L1-L4&microwave to voltage drop ratio %from delta1 and delta2 to g1 and g2
Here we evaluate g-factor of each dot. In order to convert from $\delta_{1}$ and $\delta_{2}$ to $|g_{L}|\mu_{B}B$ and $(|g_{L}-g_{R}|)\mu_{B}B$, the voltage drop ratio of three barriers, $\alpha_i$ ($i=1, 2, 3$), is required. Figure 5(a) shows $dI_{SD}/dV_{SD}$ at 12T and we mark L1-L4 for the current threshold lines from the $N=1$ Coulomb diamond. At the threshold L1, the Fermi energy of the left electrode is aligned with the energy level of the left dot. The current threshold several orders of magnitude smaller, marked by L2, indicates the onset of a cotunneling process where the right-dot energy level is aligned with the Fermi energy of the left electrode. Similarly, the Fermi level of the right electrode is aligned with the right (left) dot at the threshold line marked as line L3 (L4). The slopes of these threshold lines L1-L4, that is, -32, -54, 46 and 62, respectively, give $\alpha_i$ ($i=1, 2, 3$) of 0.27, 0.19 and 0.54, respectively, at $V_{SD}\sim$0.
Figure 5(b) shows the effect of microwave irradiation on the resonant current peak. With increasing microwave power, satellite current peaks appear on the left (right) side of the main peak, owing to interdot tunneling upon the absorption (emission) of microwaves. Such microwave-assisted tunneling in a double dot has been observed in lateral and vertical dots \cite{Oosterkamp, Kodera}. At the satellite peak position, interdot detuning is equal to the microwave photon energy. Thus the distance between the main and the satellite peak (marked as two dotted lines in Fig. 5 (b)) gives $\alpha_2$ of 0.193, which is close to the value for $V_{SD}\sim0$. Similar values are obtained for different current peaks at different $V_{SD}$'s from 0 up to 100$~$mV, indicating that $\alpha_2$ is independent of $V_{SD}$. On the other hand, voltage drop ratios for outer barriers $\alpha_i$ ($i=1,3$) are found to depend on $V_{SD}$. Indeed, in Fig. 1, the current threshold line from $N=0$ at positive $V_{SD}$ is not straight but bends upward. This indicates decreasing $\alpha_1$ with increasing $V_{SD}$. As the slope of the current threshold lines is -12, as determined from Figs. 2(a)-(d), $\alpha_1$ at $V_{SD}\sim$35$~$mV is smaller than $\alpha_1$ at $V_{SD}\sim0$ by 12/32. Thus $\alpha_i$ ($i=1,2,3$) is 0.10, 0.19 and 0.71, respectively, at $V_{SD}\sim$35$~$mV \cite{Asysmetry}. Using these $\alpha_i$'s as well as the slope at $V_{SD}\sim$35$~$mV, we obtain $|g_{L}|$ of 0.89 for our In$_{0.04}$Ga$_{0.96}$As dot, and the g-factor difference between the two dots is 0.56. Thus, $|g_{R}|$ for our GaAs dot is 0.33. The value for the GaAs dot is consistent with that obtained in the previous work on a vertical single dot with the same 10-nm-thick GaAs well \cite{Kodera}. The value of 0.89 for InGaAs seems to be large compared with the value obtained in the study of the g-factors of bulk In$_{x}$Ga$_{1-x}$As \cite{Weisbuch}, probably owing to the increasing density of In at the center of the well rather than at the InGaAs/AlGaAs interfaces, as is seen in the self-assembled InGaAs dot systems \cite{Liu}. A stress in the thin InGaAs layer can be another possible reason for large g-factor.

%phonon effect, invisible kinks in negative bias%double threshold in 14T
The width of the current peak line in both regions A and D is $\sim$0.5$~$mV in $V_{SD}$, and corresponds to $\sim$0.1$~$meV using $\alpha_2$. This width gives an upper bound for possible contributions from phonon-absorbing/emitting tunneling, and is smaller than the Zeeman energy difference in high magnetic field ($\gtrsim$10$~$T). Thus phonon absorption/emission for a misaligned spin sublevel in region B (C) in Fig. 3 is negligible. This point is also confirmed by a more detailed calculation, as in Fig. 3, where the phonon effects are included\cite{Tokura}.
As depicted in the right inset of Fig. 1, there are some unoccupied levels located a few meV's below the aligned level in the right dot. Thus, in addition to the resonant tunneling through the two aligned levels, there is always a tunneling process with meV-phonon emission down to these low-lying levels. Such meV-phonon-emitting tunneling gives a background current of the order of $\sim$10$~$pA, which defines the current threshold from the $N=0$ region, as seen in Fig. 1 and Figs. 2(a)-(d).
Resonant tunneling current peak lines are also seen for negative $V_{SD}$. However, most of the peak widths are large and no clear kink structure is observed. For negative $V_{SD}$, the direction of the $\Delta/2$ shift would be opposite, and  $\delta_{1}$ of the kink should be smaller as it is governed by the g-factor in the right dot, which is smaller for GaAs. This may be one of the reasons why we were not able to detect clear kink structures.
In Fig. 2(d), there seems to be an additional line parallel to the current threshold line from the $N=0$ region. We found that these lines, which appeared irregularly with some magnetic fields and $V_{SD}$'s, may be due to a fluctuation of the density of states in the electrode \cite{Kouwenhoven}.

%conclusion and outlook
In conclusion, we measured the single electron transport through two zero-dimensional states with different g-factors. We found that the resonant tunneling is suppressed even when one of the Zeeman sublevels is aligned within the transport window. Finite level broadening of zero-dimensional states partially relieves this bottleneck effect, and gives a current peak when level detuning is set to half the Zeeman energy difference.

%acknowledgement
The authors thank to H. Kosaka, T. Nakaoka for fruitful discussion and S. Schneider for
experimental assistance. This work was supported by the CREST-JST and RIKEN-NCTU Joint Graduate School Program.

%bibliography

\newpage %Just because of unusual number of tables stacked at end


\begin{thebibliography}{}
\bibitem{Asysmetry} Although our triple barrier structure is grown nominally symmetric against the center barrier, the electron density in the pillar structure (left electrode) decrease due to the microfabrication, causing an "offset" of interdot detuning even when $V_{SD}=0$ \cite{Vertical double dot, 0D0D, OnoSB}.Decreased electron density leads to a thicker depletion layer in the gradiently n-doped electrode, thus increasing the effective barrier thickness, which is a sum of the depletion layer at the electrode and the AlGaAs barrier. Weaker electron screening decreases the electric field inside the effective barrier. This asymmetry is further enhanced with large positive $V_{SD}$, as depicted in Fig. 1, right inset.
\end{thebibliography}
\end{document}